\documentstyle[preprint,aps]{revtex}
\begin{document}
\draft
\title{$^3$He Transport in the Sun and the Solar Neutrino Problem}
\author{Andrew Cumming{\footnote{Present address:  Department of Physics, Univ. 
of California, Berkeley, CA 94720}}}
\address{Physics Department, Trinity College \\
Cambridge University, Cambridge CB2 1TQ, United Kingdom}
\author{W.C. Haxton}
\address{Institute for Nuclear Theory, Box 351550 \\
and Department of Physics, Box 351560 \\
University of Washington, Seattle, Washington  98195-1550}
\date{\today}
\maketitle
\begin{abstract}
Recent solar neutrino experiments have shown that both $\phi(^8$B)
and the neutrino flux ratio $\phi(^7$Be)/$\phi(^8$B) are
substantially below their standard solar model values,
leading some to discount the possibility of an astrophysical
solution to the solar neutrino puzzle.  We test this conclusion
phenomenologically and find that the discrepancies can be
significantly reduced by a distinctive pattern of core mixing
on timescales characteristic of $^3$He
equilibration.
\end{abstract}
\pagebreak 
The results of the $^{37}$Cl \cite{dav1}, SAGE/GALLEX \cite{abd2}, and
Kamioka II/III \cite{suz3} experiments are consistent with an unexpected
pattern of neutrino fluxes,
\begin{eqnarray}
\phi({\mathrm {pp}})  &\sim& \phi^{SSM}({\mathrm {pp}}) \nonumber \\
\phi (^7{\mathrm {Be}}) &\sim& 0 \\
\phi(^8{\mathrm B}) &\sim& 0.4 \phi^{SSM} (^8{\mathrm B})
\nonumber
\end{eqnarray}
where $\phi^{SSM}$ denotes the standard solar model \cite{bah4} (SSM) value. 
As $\phi(^8$B) $\sim T_c^{22}$ \cite{cas5}, where $T_c$ is the solar core temperature,
the required reduction in this flux can be achieved by lowering $T_c$
to about 0.96 of the SSM value.
However such a reduction leads to an elevated $\phi(^7$Be)/$\phi(^8$B)
flux ratio because
\begin{equation}
{\phi(^7{\mathrm {Be}}) \over \phi (^8 {\mathrm B})} \sim T_c^{-10}
\eqnum{2}
\end{equation}
in contradiction to the results of Eq. (1).

This qualitative expectation - that it is very difficult to
simultaneous reduce the $^8$B neutrino flux and the $^7$Be/$^8$B
flux ratio - has been examined and confirmed in several
careful studies.  These have included variations in the
SSM parameters within generally accepted ranges \cite{bah6}, variations
far outside such ranges \cite{cas5,hat7}, and explorations of nonstandard models 
\cite{bah8}.
This has led many in the field to favor nonastrophysical solutions
to the solar neutrino problem, such as neutrino oscillations.

It is clear that no astrophysical solution of the solar neutrino
problem will give a perfect fit to the results of existing
experiments: the measurements are inconsistent with any
combination of undistorted $^8$B, $^7$Be, and pp neutrino fluxes at
a confidence level of about 2$\sigma$ \cite{ber9}.  Yet it is also
clear that a compelling argument for a resolution in terms
of new particle physics must rest on the more dramatic
discrepancy (often estimated at 5$\sigma$) that exists between experiment and the
flux predictions of standard and nonstandard models.  Thus it
is important to determine whether a nonstandard model
might exist in which the naive $T_c$ dependence described above
is circumvented.

If such a model exists, the associated physics could be subtle.  For this
reason we will try a simple-minded approach - changing the SSM
phenomenologically - putting aside for the moment the deeper issue of
the underlying mechanism.  We consider perturbations of the Bahcall-Pinsonneault (BP) SSM
(without He or metal diffusion), constrained by three
conditions.  First, we retain all of the standard nuclear
and atomic microphysics, e.g., nuclear cross sections and
opacities.  This reflects our view that SSM ``best values"
and uncertainties are sensible chosen, and our reluctance to
produce a trivial solution to the solar neutrino problem
by $ad \,\,hoc$ adjustments of parameters.  Second, we require
that our phenomenological changes not alter the known solar
luminosity.  In a steady-state model this places the following
constraint on the neutrino fluxes
\begin{equation}
6.481 \cdot 10^{10}/{\mathrm {cm^2 sec}} = \phi {\mathrm {(pp)}} + 0.9561 \, 
\phi \, (^7 {\mathrm 
{Be}}) + 0.5075 \, \phi \, (^8 {\mathrm B})
\eqnum{3}
\end{equation}
where we ignore small changes due to variations in the CNO cycle
contribution.  In our calculations this constraint was enforced
in a somewhat crude fashion, by retaining the BP temperature
profile of Table VII of Ref. \cite{bah4} but allowing an overall
rescaling.  Third, we require
the model to be steady-state, demanding where
appropriate equilibrium in the production and consumption
of pp chain ``catalysts" such as D, $^3$He, and $^7$Be.

The third condition is typically implemented locally
in the SSM, while the weaker condition of global equilibrium
is still compatible with a steady-state sun.  Thus these
conditions allow a broad class of continuously mixed suns
where pp chain products ($^4$He as well as the nuclei mentioned
above) are transported.  The possibility of slow mixing,
which would allow the ``catalysts" to remain at their local
equilibrium values but tend to homogenize the H and $^4$He
in the mixed portion of the core, was explored many years ago \cite{eze10,sha11}
and is known not to produce the flux pattern of Eq. (1).
Likewise it can be reasonably argued that the transport of
D and $^7$Be is less interesting because of the requirement
of very large mixing
velocities: D is destroyed almost instantaneously, while the
solar lifetime of $^7$Be is about 100 days.  On the other hand,
$^3$He is intriguing.  It is produced in the pp chain by
p + p $\to$ D + e$^+$ + $\nu$ followed by D + p $\to ^3$He
+ $\gamma$ at a rate
\begin{equation}
r_{11} \propto X^2_1 \,\, T_7^4
\eqnum{4}
\end{equation}
where $X_1$ and $T_7$ are the local mass abundance of hydrogen
and temperature (units of 10$^7$ K).  It is consumed by the
competing reactions
\begin{equation}
^3{\mathrm {He}} + {^3 {\mathrm {He}}} \to {^4 {\mathrm {He}}} + 2 {\mathrm p}
\eqnum{5a}
\end{equation}
\begin{equation}
^3{\mathrm {He}} + {^4 {\mathrm {He}}} \to {^7 {\mathrm {Be}}} + \gamma
\eqnum{5b}
\end{equation}
with the former being dominant.  As its rate is
\begin{equation}
r_{33} \propto X^2_3 \,\, T_7^{16} \eqnum{6}
\end{equation}
where $X_3$ is the abundance of $^3$He,
it follows that, in local equilibrium,
\begin{equation}
X_3 \sim 7 \cdot 10^{-4}\, X_1 \,\, T_7^{-6} \eqnum{7}
\end{equation}
Thus the SSM $^3$He equilibrium abundance increases sharply with
radius (decreasing $T_7$), as does the time required to reach equilibrium,
which varies approximately as $T_7^{-10}$. For example, the time required to 
reach 99\% of the $^3$He equilibrium value at  
 r $\sim$  0.1 R$_{\odot}$ is $\sim 5
\cdot 10^6$
years.  The SSM \cite{bah4} predicts that today's sun has reached $^3$He
equilibrium for r $\alt$ 0.27 R$_{\odot}$.

We introduced changes in the SSM equilibrium $^3$He profile, constrained
by the requirement of $^3$He global equilibrium in the core.
Such changes alter the competition between the ppI, ppII, and
ppIII cycles and thus affect the
luminosity.  To recover the correct luminosity we adjusted
the overall scale of the BP temperature profile.  This procedure must
then be iterated to 
convergence.  As $^3$He mixing timescales are short compared
to overall solar evolution, H and $^4$He were assumed to be homogeneous
throughout the mixed portion of the core.
We chose very simple, piecewise constant $^3$He profiles, as
our goal was to determine the qualitative features of any
$^3$He distribution consistent with Eq. (1).

Profiles that simultaneously produced a reduced $\phi(^8$B)
(we selected values near 0.4 of the SSM) and a
reduced flux ratio $\phi(^7$Be)/$\phi(^8$B) had
a characteristic shape: an order-of-magnitude elevation in the $^3$He abundance, relative
to the equilibrium value, at small radii, and a depletion
at large r.  The breadth and height of the region of elevated
abundance can be adjusted over some range.  The corresponding
temperature scale factors ranged from near 1 to about 0.93,
with 0.95 being a typical value.  Thus the resulting sun is a
cooler one, consistent with the increase in ppI terminations
demanded by Eq. (1).  Some typical results are given in
Table I and illustrated in Fig. 1.  (Fig. 1 is, of course, a
caricature: only the qualitative aspects of the $^3$He distribution
have significance.)

It is readily seen why such a change moves the neutrino flux
predictions towards the results of Eq. (1).  First, 
a large fraction of the produced $^3$He is burned out of
equilibrium at small r.  The ppI terminations are governed
by reaction (5a), which is quadratic in the $^3$He abundance,
while the competing reaction (5b) is linear.  Thus the
rate of ppII+ppIII terminations relative to ppI terminations
is reduced in direct proportion to the $^3$He excess,
suppressing both the $^7$Be and $^8$B neutrino fluxes.
However, when reaction (5b) does occur, short-lived $^7$Be is
produced at small r, where the ambient temperature is high.
This favors ppIII terminations over ppII terminations, leading
to a suppressed $\phi(^7$Be)/$\phi(^8$B) flux ratio.
The combined effects of the reduced (ppII+ppIII)/ppI
and enhanced ppIII/ppII branching ratios yield a somewhat reduced
$^8$B neutrino flux and a significantly reduced $^7$Be flux.

Such a pattern of $^3$He burning can only arise if there is 
core mixing on a timescale characteristic of $^3$He equilibration.
In fact, the profile of Fig. 1 suggests a rather specific mixing mechanism.
First, there must be a relatively rapid downward flow of 
$^3$He-rich material from large r; the speed must be 
sufficient to take a mass element well past the 
usual equilibrium point, into a region where the rapidly 
decreasing local lifetime of $^3$He final results in sudden $^3$He
ignition.  This mass element, now depleted in $^3$He and
buoyant because of the energy release, must return to
large r sufficiently slowly to allow the p+p reaction to
replenish the $^3$He.  This flow is depicted in Fig. 2.  As
we are assuming a steady-state process in which any mass
element is roughly equivalent to any other, 
each mass element must, on average, remain within
a radial shell bounded by r and r+dr for a time proportional
to the mass dM(r) contained within that shell.  This condition
would be satisfied if the slow upward flow is broad
with a local velocity inversely proportional to dM(r) -
the kind of flow that would result from displacement from below.
Such upward flow will produce a positive $^3$He gradient,
as in the SSM; but the upward flow must be sufficiently fast to
keep the $^3$He below its local equilibrium value to prevent
burning at large r.
To keep the circulation steady, the rapid downward flow clearly
must be localized, e.g., perhaps in narrow plumes.

We would like to stress that we are not proposing this as a
solution to the solar neutrino puzzle.  But we are suggesting
that arguments against an astrophysical solution based on the
naive $T_c$ dependence of neutrino fluxes are likely
overstated.  We have sketched how the naive
expectations might be circumvented by core mixing.

Yet there are amusing aspects of the mixing that we
would like to explore further, with the understanding that our
comments are quite speculative:

1)  Although core mixing on the timescale for $^3$He equilibration
has been considered previously, including in the
early work of Shaviv and Salpeter \cite{sha11}, we believe the
possibility of different upward and downward flow velocities has
not been explored.  In Table I we give estimates of these velocities
for a various profiles of the type illustrated in Fig. 1.
The downward plume velocity (taken to be constant) is 
fixed by the condition that a mass element with the necessary
$^3$He abundance will be swept to the appropriate point before
burning commences.  Defining the onset of burning as a depletion 
of the $^3$He to 80\% of its initial value at large r, we find
transport times $\tau_{\downarrow}$ = (2-12) $\times$ 10$^6$ years,
or velocities on the order of 10-100 m/y.

The temperature and volume of the mass element will increase as
$^3$He burning proceeds under the condition of constant pressure,
resulting in an upward acceleration due to the buoyancy.  We lack
a sufficiently detailed physical picture to model this: clearly
the temperature trajectory will depend on a competition between
energy generation and thermal transport, with the later depending
on the plume geometry.  Thus we have depicted this part of the
trajectory in Fig. 2 by a dashed line.  Qualitatively the rising
temperature will increase the suddenness of the $^3$He burning 
and further suppress the ppII/ppIII branching ratio, relative
to the estimates of Table I.

As the rising mass element is now depleted in $^3$He, when it 
cools it should be similar to, and merge with, the surroundings.
We envision the subsequent upward flow as described above -
slow and global, proportional to dM(r) - and have checked whether
the necessary accumulation of $^3$He could then occur, given 
the constraint that the produced $^3$He not burn at large r.
The results are somewhat interesting.  For 
profiles similar to Fig. 1 in which the mixing was confined to
the inner core, r $\lesssim$ 0.2 R$_{\odot}$, this can be
achieved only if the slow, upward flow (the solid part of the
upward trajectory in Fig. 2), begins at relatively large r.
It then becomes complicated to explain how the overall flow
can be viewed as one where every mass element within the mixed core cycles in an
equivalent way.  On the other hand, if the mixed region extends
to large r, 0.25-0.30 R$_{\odot}$, the slow upward flow
dominates the mixed core, beginning typically at r $\sim$
0.1 R$_{\odot}$.  The results in Table I are of this class,
with the starting point for the upward flow defined by the requirement
that at least 80\% of the produced $^3$He survives unburned.  The
corresponding times required for the flow fall in the
range $\tau_{\uparrow}$ = (4.2-15.2) $\times$ 10$^7$ years, roughly an order
of magnitude longer than the corresponding $\tau_{\downarrow}$.

We are somewhat surprised that simple flow patterns could
qualitatively produce the $^3$He burning profile depicted in
Fig. 1, as the latter was deduced from Eqs. (1) phenomenologically
and without regard for physical plausibility.  In fact, the
resulting preference for larger mixed cores, ones encompassing
most of the region where the SSM $^3$He gradient has been
established, is a rather pleasing result.  Of course, our 
inability to model the small r region where $^3$He is burned is
an important caveat.

2) The possibility of flows in the background of a positive
$^3$He gradient raises the old issue of the ``solar spoon"
instability first discussed by Dilke and Gough \cite{dil12}.  These authors
pointed out that the SSM is unstable to large amplitude perturbations
because the energy released by enhanced $^3$He burning can exceed
the work against gravity required to force a mass element at 
large r through the denser material below.
The solar spoon has been most often discussed 
as a trigger for periodic core mixing \cite{mer13}.

In the case of the continuous flow postulated in 1), the core
would remain homogenized in H and $^4$He while still permitting
a $^3$He gradient, an amusing variation on the solar spoon.
The plume flow we have described would then be essentially
adiabatic.  Large-scale adiabatic flow that would allow the sun
to produce the required luminosity more efficiently (i.e., by
burning at a lower temperature) has a certain attractiveness.
On the other hand, one clearly needs to explain how the plumes maintain
their chemical identity as they descend through the $^3$He-poor
surroundings.

Speculations about a persistent convective core \cite{rox14}
could be relevant to the question of how the plumes are first
generated.  The core of the early sun is
believed to be convectively unstable prior to the establishment
of equilibrium in the pp and CNO cycles: $\eta$ = dlog$\epsilon$/dlogT,
where $\epsilon$ is the energy generation rate, is initially
in excess of the critical value of about 5.0.
While the initial instability is due to
the out-of-equilibrium burning of $^{12}$C to $^{14}$N, Roxburg \cite{rox14} has
suggested that $^3$He transport by convective overshooting might
help to maintain the conditions for convection up to present times.
We suspect that the flow postulated in 1) would be
convectively unstable in the region where the $^3$He is being
burned, as the disequilibrium is so similar to the early sun.
The fact that the flow is both driven by and maintains the 
disequilibrium opens up the possibility of persistent mixing:
if one introduces a perturbation to speed the mixing, this
should diminish the $^3$He gradient and energy production,
thus slowing the mixing.  That is, the mixing, if otherwise 
viable, could prove to be stable to perturbations.

3) Such a mixed core will have other astrophysical consequences.
For example, galactic evolution models \cite{gal15,dea16} predict
$^3$He abundances in the presolar nebula and in
the present interstellar medium (ISM) that are substantially (i.e., a
factor of five or more) in excess of the observationally
inferred values.  This enrichment of the ISM is driven by low-mass
stars in the red giant phase, when the convective envelope reaches
a sufficient depth to mix the $^3$He peak, established during
the main sequence, over the outer portions of the star.  The
$^3$He is then carried into the ISM by the red giant wind.
This difficulty prompted Galli et al. \cite{gal15} to suggest an enhanced
$^3$He+$^3$He cross section, which would suppress the main sequence
$^3$He peak, as the solution most compatible with observation.
While the galactic evolution of $^3$He is clearly a complex
problem, it is interesting that the mixing we have discussed lowers 
the main sequence $^3$He abundance at large r.

The envisioned mixing will also change thermal and composition
gradients and thus the core sound speed,  
affecting the helioseismology \cite{gou17}.  If helioseismology
can rule out the postulated mixing,
it would suggest that astrophysical solutions to the solar neutrino
problem would have to be more exotic than the steady-state models
considered here.

As the core mixing lowers $T_c$ and enhances ppI burning, it 
should slow stellar evolution somewhat.
However, it is generally believed \cite{sha11} that mixing
confined to the core (r $\lesssim$ 0.4R$_{\odot}$) will not affect 
main sequence and red giant evolution to
a degree that would be apparent in the color-magnitude diagram.

In summary we have argued that the naive $T_c$ dependence
of solar neutrino fluxes could be circumvented in models
where $^3$He is transported into the core.  Thus the $T_c$ argument
by itself is not sufficient to rule out an astrophysical
solution to the solar neutrino problem, even if one limits
the discussion to steady-state models with conventional
microphysics.  We then pointed out that the $^3$He profile
consistent with Eq. (1) is suggestive of a rather unusual steady-state
mixing pattern involving rapid filamental flow downward and
a slow, broad restoring flow upward.  Whether such flow could
occur in the sun is entirely speculative, and the consistency
of the resulting solar model with
helioseismology is an open question.  Some of the
issues raised by the hypothesized mixing are
reminiscent of such ``closet skeletons"
as the SSM $^3$He instability, an early convective core, and
galactic $^3$He evolution.

This work was supported by the University of Washington
and National Science Foundation Research Experiences for
Undergraduates program (AC) and by the US Department of
Energy (WH).

\pagebreak

\begin{table}
\caption{Modified $^3$He profiles.  The inner
(enhanced $^3$He) and outer (depleted) portions of the mixed core are
denoted by $\Delta$r$_{\mathrm {I}}$ and $\Delta$r$_{\mathrm {O}}$ and specified in units 
of R$_{\odot}$.  The third column gives both the absolute and 
normalized (relative to equilibrium, in parentheses) $^3$He mass
fractions in the inner core.  The temperature $T_c$, $\phi(^8$B),
and $\phi(^7$Be) results are normalized to SSM values.  X$_3^{eq}$
is the equilibrium $^3$He mass fraction at the outer edge of the
mixed core. $\tau_{\downarrow}$ and $\tau_{\uparrow}$ are the transit
times for sinking and rising mass elements (see text).}
\begin{tabular}{lllllllll}
$\Delta$r$_{\mathrm {I}}$&$\Delta$r$_{\mathrm {O}}$&
X$_3^{\mathrm {I}}$ (10$^{-3}$)&
$T_c$&$\phi(^8$B)&$\phi(^7$Be)&
X$_3^{eq}$ (10$^{-3}$)&
$\tau_{\downarrow}$ (10$^7$ y)&
$\tau_{\uparrow}$ (10$^7$ y) \\ \hline
0.000-0.020&0.020-0.23&0.334 (12.3)&0.949&0.40&0.24&1.62&0.21&14.9 \\
0.000-0.031&0.031-0.31&0.219 (6.3)&0.932&0.40&0.33&13.3&0.85&4.2 \\
0.010-0.025&0.025-0.26&0.269 (8.7)&0.940&0.40&0.29&4.01&0.43&7.0 \\
0.020-0.025&0.025-0.22&0.329 (12.0)&0.953&0.41&0.26&1.32&0.28&15.2 \\
0.020-0.031&0.031-0.27&0.246 (7.6)&0.940&0.41&0.32&4.74&0.64&5.7 \\
0.031-0.035&0.035-0.24&0.305 (10.0)&0.952&0.40&0.31&1.88&0.52&11.3 \\
0.039-0.045&0.045-0.27&0.250 (7.0)&0.948&0.40&0.39&4.44&1.16&6.0 \\

\end{tabular}
\end{table}

\pagebreak

\pagebreak

\begin{figure}
\caption{The dashed line gives the SSM equilibrium $^3$He mass fraction
rescaled for a core temperature $T_c$=0.940 $T_c^{SSM}$.   
For the same temperature, the solid line is a modified $^3$He profile 
producing an equivalent $^3$He burning rate, 
the correct luminosity, and the neutrino fluxes listed in the
third row of Table I.}
\label{fig1}
\end{figure}

\begin{figure}
\caption{A schematic of the circulation of a mass element that 
would produce a $^3$He burning profile qualitatively similar to Fig. 1.  The portion of the solid
line with the downward arrow represents a $^3$He-rich plume
descending toward the core; the remainder represents the slow
upward flow resulting from displacement from below.  
The dashed line represents the process of
$^3$He ignition, buoyancy, and subsequent cooling.  This
portion of the circulation has not been modeled numerically.}
\label{fig2}
\end{figure}

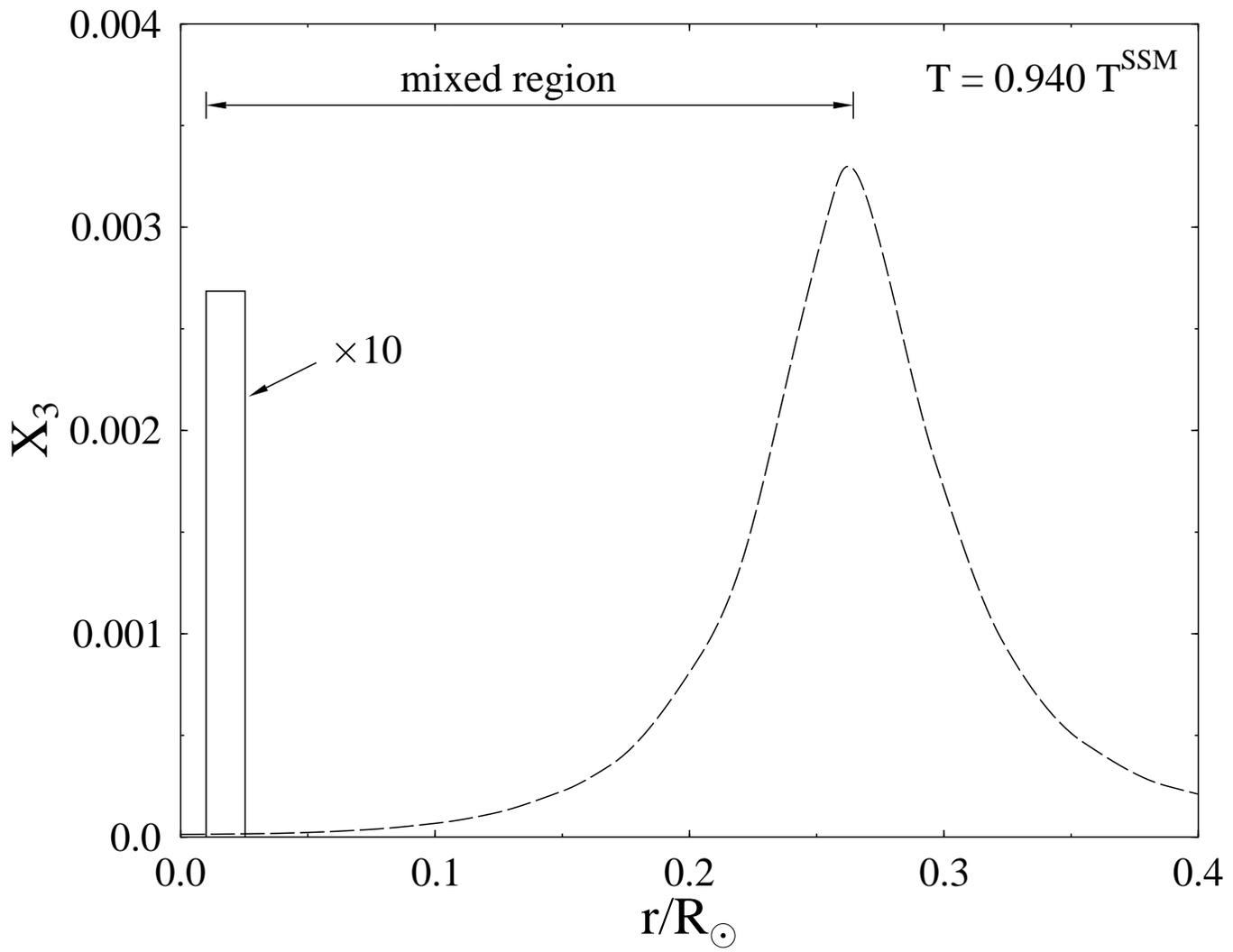
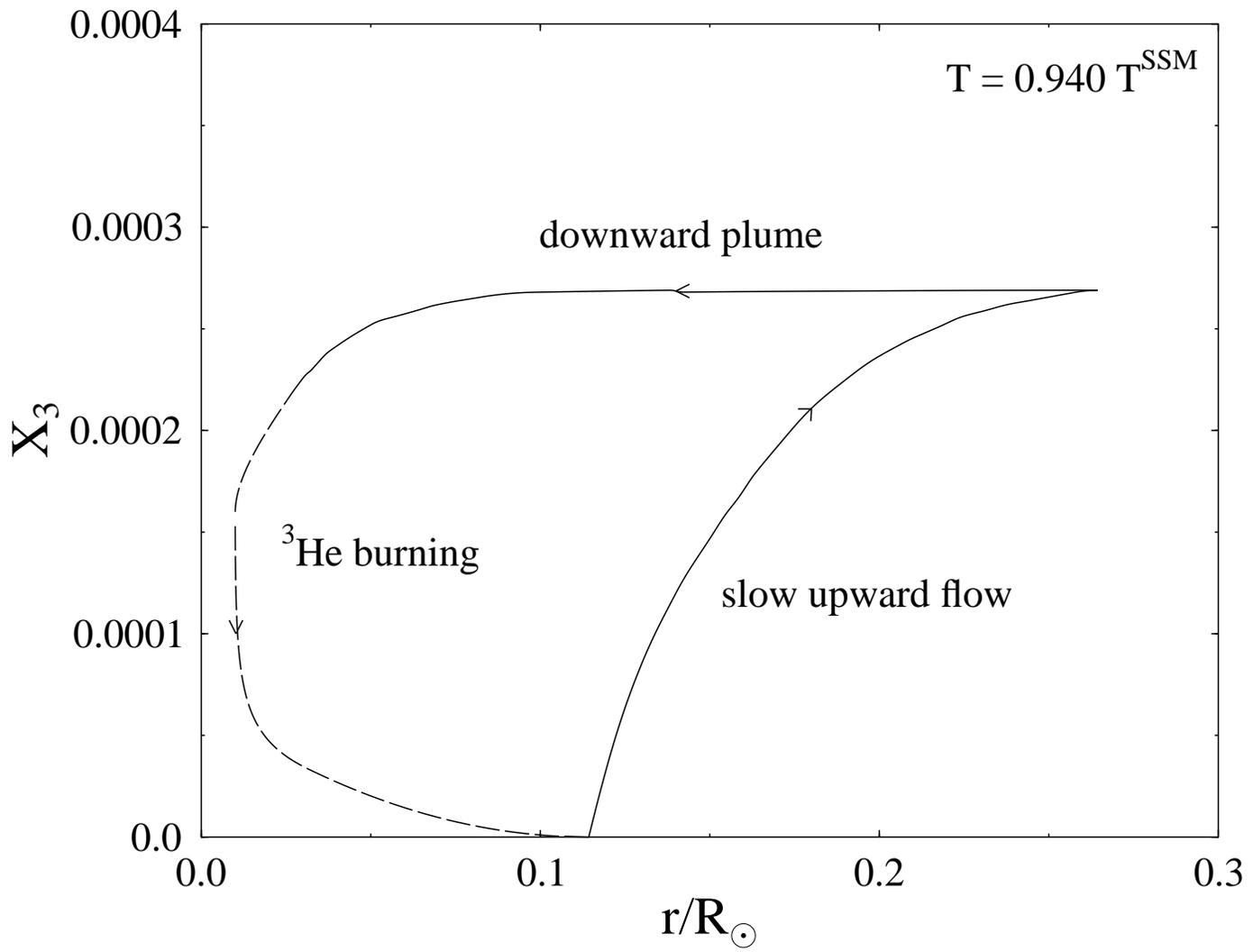
\end{document}